\documentclass{PoS}

\title{VERITAS Search for Magnetically-Broadened Emission From Blazars}

\ShortTitle{VERITAS Search for Magnetically-Broadened Emission From Blazars}

\author{\speaker{Elisa Pueschel} for the VERITAS Collaboration\\
        University College Dublin, Dublin, Ireland \\
        E-mail: \email{elisa.pueschel@ucd.ie}}

\abstract{A non-zero intergalactic magnetic field (IGMF) would potentially produce detectable effects on cascade emission from blazars. Depending on the strength of the IGMF, the cascade emission may be time delayed or angularly broadened compared to the blazar's primary, unscattered emission. Ground-based imaging atmospheric-Cherenkov telescopes, such as VERITAS, have the precise angular resolution needed to search for magnetically-broadened emission. We present the latest VERITAS results on the search for extended gamma-ray emission, based on observations of a number of strongly-detected TeV blazars at a range of redshifts. The consequent constraints on the strength of the IGMF are discussed.
}

\FullConference{The 34th International Cosmic Ray Conference,\\
		30 July- 6 August, 2015\\
		The Hague, The Netherlands}

\begin{document}

\section{Introduction}
The intergalactic magnetic field (IGMF) is a weak, possibly primordial field that permeates the voids between galaxies~\cite{Durrer}. One method to probe the strength of the IGMF is to use observations of distant blazars with imaging atmospheric Cherenkov telescopes (IACTs). A fraction of the TeV-range $\gamma$-rays emitted by blazars will undergo $e^{+}e^{-}$ pair production upon interaction with the extragalactic background light (EBL). These secondary particles will then interact via inverse Compton scattering with low-energy cosmic microwave background (CMB) photons, up-scattering the photons to higher energies. To illustrate the energies involved, a $\sim$10~TeV photon emitted by a blazar will be reprocessed to secondary cascade photons of $\sim$100~GeV energy~\cite{Elyiv}. Thus, it is possible that if a blazar has a sufficiently hard spectrum and sufficiently high energy cut-off, cascade emission could be observed by an IACT arrays such as VERITAS. The currently operating IACTs have energy thresholds around 100~GeV, thus for an appreciable fraction of the emission above 100~GeV to be due to cascade emission, the source must have a cut-off at TeV energies.

In the presence of a non-zero IGMF, the cascade emission will be angularly broadened and time delayed as the deflection of the intermediate electrons and positrons by the magnetic field increases the path length between the source and the observer compared to unscattered emission. The magnitude of the angular broadening and time delay varies with the IGMF strength and can be divided into three regimes. For 10$^{-12}$~G~$\lesssim$~B~$<$~10$^{-7}$~G, the electron-position pairs will be isotropized in the vicinity of the blazar, forming a pair halo detectable as an angular broadening compared to unscattered emission and the IACT's point spread function (PSF)~\cite{Aharonian}. For 10$^{-16}$~G~$<$~B~$\lesssim$~10$^{-12}$~G, the bulk of the $e^{+}e^{-}$ pairs will not be isotropized. The cascade will be slightly time delayed as well as angularly broadened compared to unscattered emission~\cite{Elyiv}. For B~$<10^{-16}$~G, the predicted angular broadening is too small to be disentangled from the PSF, at least for the angular resolution of the currently operating IACTs, and the cascade can only be detected via observation of a time delay compared to unscattered emission~\cite{Takahashi}. This study focuses on searching for angular broadening rather than time delays, and is thus insensitive to the third regime of IGMF strengths. 

In addition to the dependence on the intrinsic source spectrum, the projected sensitivity to cascade emission depends on the source redshift, due to the evolution of the EBL with redshift. For very distant sources (z~$\gtrsim$~0.2), the cascade emission is too compact to be distinguishable from the beamed emission, whereas for nearby (z~$\lesssim$~0.1) sources, the cascade emission is too broad to be distinguishable from the cosmic-ray background.

Previous searches for extended emission around blazar have been performed by MAGIC (using Mrk 501 and Mrk 421), $Fermi$-LAT (using a large blazar sample), VERITAS (using Mrk 421) and H.E.S.S. (using 1ES 1101-232, 1ES 0229+200 and PKS 2155-304)~\cite{MAGIC, Fermi-LAT, Mrk421, HESS}. 

\subsection{Source selection}
The sources used in this analysis were selected for optimal sensitivity to magnetically-broadened emission. As stated above, the predicted cascade fraction is greatest for hard-spectrum sources with high energy cut-offs. The spectral index $\Gamma$ measured by $Fermi$-LAT in the GeV band is used to approximate the intrinsic index~\cite{2FGL}. Sources with $\Gamma$<2 were selected. Note that in the absence of a $Fermi$-LAT-measured spectral index for 1ES 0229+200, an intrinsic spectral index $\Gamma$=1.5 is assumed. The cut-offs in the intrinsic spectrum were taken from the joint GeV-TeV fits performed in~\cite{Cutoffs}. For sources not appearing in ~\cite{Cutoffs} or consistent with no spectral break, a cut-off energy of 10 TeV assumed. This value was selected based on its position at the lower edge of the range of cut-off energies that could still yield a spectrum consistent with no spectral break.

To minimize statistical uncertainties, strongly-detected sources were used, ideally those with $>$10$\sigma$ significance. This requirement was softened for sources falling in the redshift range with the greatest expected sensitivity (z$\simeq$ 0.1-0.2). Although sources at the optimal redshift were preferentially selected, sources at higher and lower redshifts were also included in the analysis. In the case of a detection of a broadening of the emission, this would allow tests of the predicted redshift dependence. The final source list, together with the assumed intrinsic spectral index, energy cut-off, redshift and detection significance in the data sample used is shown in Table~\ref{sourcelist}. 

\begin{table}
\centerline{
\begin{tabular}{ccccc}
 & $z$ & $\Gamma$ & $E_{cut} [TeV]$   & $\sigma_{detect}$   \\
\hline
\hline
Mrk 421 & 0.031 & 1.771$\pm$0.012 & 0.692 & 185.3 \\
Mrk 501 & 0.034 & 1.738$\pm$0.027 & 1.630 & 94.8 \\
H 1426+428 & 0.129 & 1.316$\pm$0.123 & 0.570 & 7.6 \\
VER J0521+211 & 0.108 & 1.928$\pm$0.034 & 10 &  23.2 \\
1ES 0229+200 & 0.14 & 1.5 & 10 & 10.3 \\
1ES 1218+304 & 0.182 &  1.709$\pm$0.067 & 1.010 & 35.5\\
PG 1553+113 & 0.5 & 1.665$\pm$0.022 & 10 & 46.0 \\
\hline
\hline
\end{tabular}
}
\caption{Source list, with the redshift, the assumed intrinsic spectral index (based on the $Fermi$-LAT measured index), high energy cut-off, and detection significance in the dataset used for this analysis.}
\label{sourcelist}
\end{table}

\section{VERITAS observations}
VERITAS is an array of four 12 meter IACTs~\cite{VTS} located at Fred Lawrence Whipple Observatory in southern Arizona. VERITAS is sensitive to very high energy $\gamma$-rays in an energy range from 85 GeV to greater than 30 TeV. VERITAS has an energy resolution of 15\% and an angular resolution of 0.1$^{\circ}$ at 1 TeV.

The data used in this analysis were collected between 2009 and 2012, prior to a major camera upgrade. As obtaining the best possible angular resolution is critical for this measurement, only runs with all four telescopes operating were used. Additionally, only runs with zenith angle of observations $<$30$^{\circ}$ were used, both because the angular resolution is better and because the energy threshold is lower than for large zenith angle observations. During strong flares, it is expected that the unscattered emission will be brighter than the cascade emission, decreasing the overall sensitivity to cascade emission. Thus, periods of high source activity were removed from the datasets. 

\section{Analysis}
All data were processed using the standard VERITAS calibration and shower parameterization methods. The $\gamma$-hadron separation was achieved using selection on mean-scaled Hillas parameters for all sources other than H 1426+428, 1ES 0229+200, and VER J0521+211. For these three sources, which were not detected to a high significance with standard $\gamma$-hadron separation, a boosted decision tree (BDT) analysis was used. The BDT analysis was not used for the other sources because the energy threshold is slightly higher for the BDT analysis, and preserving the lowest possible energy threshold was prioritized for sources already strongly detected with the standard $\gamma$-hadron separation.
 
To achieve a good angular resolution while retaining as many photon events as possible, events with images recorded in only two telescopes were discarded, while events with three and four images were retained. The angular resolution is additionally correlated with the distance between the shower core and the array center. The selection on this parameter was optimized to improve the angular resolution, resulting in an optimal value $CoreDist<$215m.

To search for extended emission, the angular distribution of a source can be compared against a known point source whose angular extent is due only to the instrument PSF. The parameter $\theta^{2}$, defined as the angular distance between the air shower's arrival direction and source's estimated location, characterizes the angular profile. The VERITAS PSF varies with energy, the azimuthal and zenith angles of observations, and the night sky background level. Thus, it was necessary to simulate a point source for each blazar studied that matched the source characteristics and observing conditions. The range and mean value of the night sky background level was matched between each source and its simulated point source. The simulations were weighted event-by-event to match their data counterparts' energy and azimuthal angular distributions. It was not possible to exactly match the sources' zenith angular distributions, as simulations were generated at discrete values for the zenith angle of observation. A zenith angle of 20$^{\circ}$ provided the best approximation of the zenith distribution in data, and was therefore used. A correction (described below) accounts for this simplification.

\begin{figure}
\centerline{\includegraphics[width=0.35\textwidth]{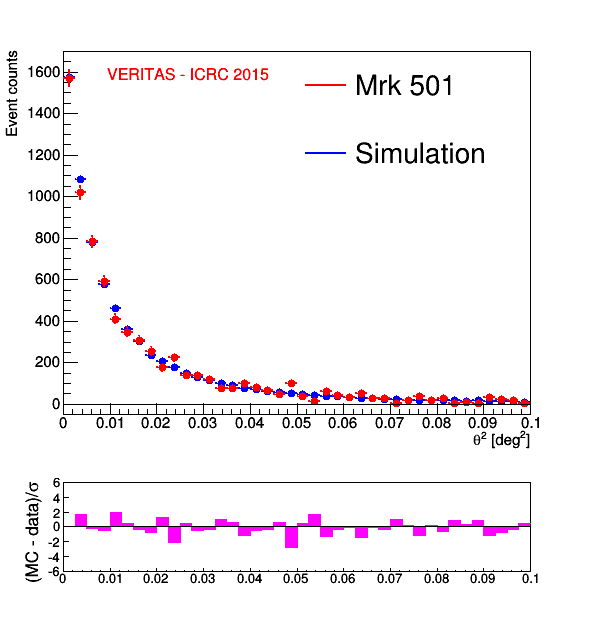}
\includegraphics[width=0.35\textwidth]{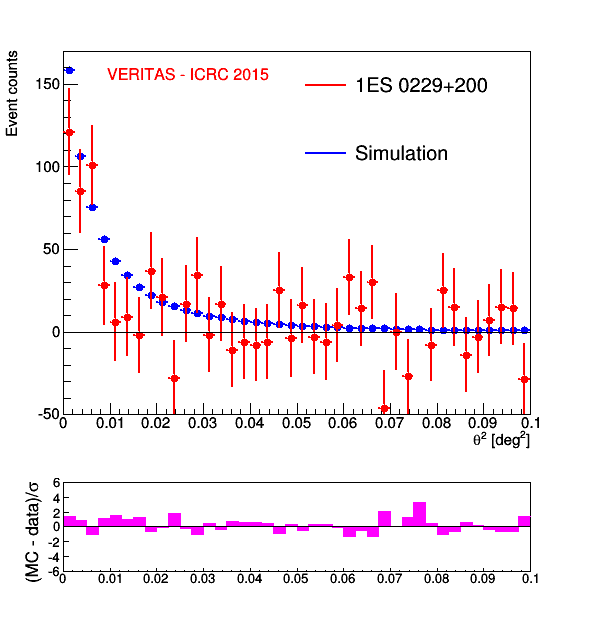}
\includegraphics[width=0.35\textwidth]{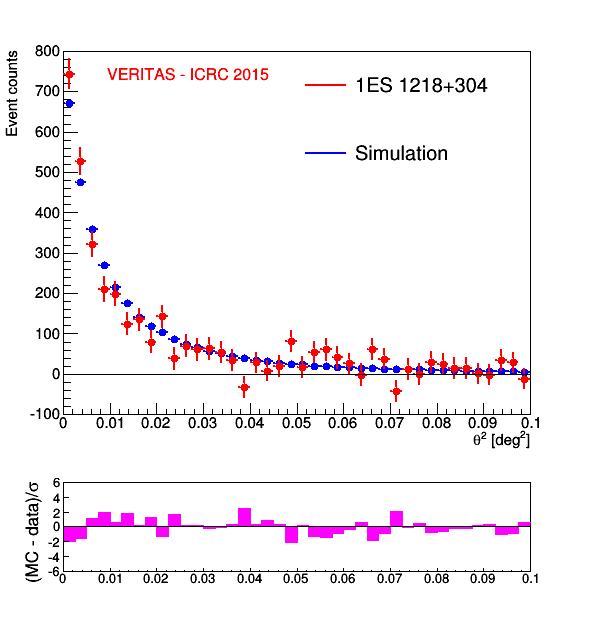}}
\caption{\small{Comparison between the angular profile of three sources and their simulated point sources. Good agreement is clear from the residual distributions. The results of a $\chi^{2}$ probability test are shown in Tab.~\protect\ref{chi2tests} for all sources.}}\label{dataMCoverlay}
\end{figure}

The agreement between the $\theta^{2}$ distributions for the sources and their matching simulated point sources was assessed first based on the derived residual distributions and a $\chi^{2}$ probability test. Examples of several overlaid data/simulation distributions are shown in Figure~\ref{dataMCoverlay}. To produce a measure of the agreement that includes systematic uncertainties and the zenith angle correction, the angular distributions were fit and the width parameters $w_{data}$ and $w_{sim}$ were compared. 

\subsection{Systematic uncertainties}
Although the dominant uncertainties in this analysis are statistical, the uncertainties introduced by mispointing and energy resolution on the simulated point sources were assessed. The uncertainty on the telescope pointing is $\sim$25$''$, which translates to an uncertainty of 0.0005$^{\circ}$ on the $\theta^{2}$ distribution.

The PSF depends on energy, thus the finite energy resolution of VERITAS introduces a further systematic uncertainty on the $\theta^{2}$ distribution. An average energy resolution of 20\% was assumed. The propagated uncertainty varies depending on the steepness of the source spectrum, and was thus calculated for each source. The uncertainties vary between 0.0005$^{\circ}$ and 0.0014$^{\circ}$.

As mentioned previously, the simulated sources were generated assuming a zenith angle of observations of 20$^{\circ}$, which only approximates the observed zenith distribution in data. This assumption was addressed by deriving a function $PSF(Ze)$ for the PSF as a function of zenith angle using a large sample of Crab Nebula observations. A PSF width $w_{Ze(obs)}$ was then derived based on the observed zenith distribution, and the difference ($w_{Ze(obs)} - w_{Ze=20^{\circ}}$) was used to correct the simulated source's fitted width. The size of the zenith corrections range from 0.0009$^{\circ}$ to 0.0023$^{\circ}$. 

\section{Results and Discussion}
The fitted widths derived from the source and simulated point source $\theta^{2}$ distributions are shown in Tab.~\ref{fitwidths}. Although some marginal tension is present - Mrk 501 appears to be broader than a point source at the 1.95$\sigma$ level and 1ES 1218+304 appears to be narrower at the 1.87$\sigma$ level - there is no significant discrepancy between the sources and their simulated point sources.

Checks of the agreement using the $\theta^{2}$ histograms based on residual distributions and $\chi^{2}$ probability tests also indicated good agreement between the sources and their simulated point sources. The p-values corresponding to the $\chi^{2}$ probability tests are shown in Tab.~\ref{chi2tests}. Only PG 1553+113 has a marginal p-value, which does not account for the zenith correction or systematic uncertainties.

\begin{table}
\centerline{
\begin{tabular}{ccc}
 & $w_{data} \pm \sigma_{stat}$ & $w_{sim} \pm \sigma_{stat} \pm \sigma_{syst} $  \\
\hline
\hline
Mrk 421 & 0.0483$^{\circ}$ $\pm$ 0.0004$^{\circ}$ & 0.0484$^{\circ}$ $\pm$ 0.0002$^{\circ}$ $\pm$ 0.0009$^{\circ}$ \\
Mrk 501 & 0.0503$^{\circ}$ $\pm$ 0.0008$^{\circ}$ & 0.0481$^{\circ}$ $\pm$ 0.0003$^{\circ}$  $\pm$ 0.0007$^{\circ}$\\
1ES 0229+200 & 0.0429$^{\circ}$ $\pm$ 0.0075$^{\circ}$ & 0.0461$^{\circ}$ $\pm$ 0.0003$^{\circ}$  $\pm$ 0.0011$^{\circ}$\\
H 1426+428 & 0.0591$^{\circ}$ $\pm$ 0.0109$^{\circ}$ & 0.0547$^{\circ}$ $\pm$ 0.0003$^{\circ}$ $\pm$ 0.0014$^{\circ}$\\ 
VER J0521+211 & 0.0428$^{\circ}$ $\pm$ 0.0004$^{\circ}$ & 0.0451$^{\circ}$ $\pm$ 0.0002$^{\circ}$ $\pm$ 0.0012$^{\circ}$\\ 
1ES 1218+304 & 0.0468$^{\circ}$ $\pm$ 0.0020$^{\circ}$ & 0.0507$^{\circ}$ $\pm$ 0.0003$^{\circ}$ $\pm$ 0.0007$^{\circ}$ \\ 
PG 1553+113 & 0.0558$^{\circ}$ $\pm$ 0.0021$^{\circ}$ & 0.0521$^{\circ}$ $\pm$ 0.0002$^{\circ}$ $\pm$ 0.0011$^{\circ}$ \\ 
\hline
\hline
\end{tabular}
}
\caption{Fitted width of the $\theta^{2}$ distribution for sources and their corresponding simulated point sources.}
\label{fitwidths}
\end{table}

\begin{table}
\centerline{
\begin{tabular}{cc}
  & $p-value$   \\
\hline
\hline
Mrk 421 & 0.19 \\
Mrk 501 & 0.38  \\
1ES 0229+200 & 0.30 \\
H 1426+428 & 0.95  \\ 
VER J0521+211 & 0.31 \\ 
1ES 1218+304 & 0.52 \\ 
PG 1553+113 & 0.003 \\ 
\hline
\hline
\end{tabular}
}
\caption{Results of a comparison of the $\theta^{2}$ histograms for the sources and their simulated point sources with a $\chi^{2}$ probability test.}
\label{chi2tests}
\end{table}

\subsection{Model-dependent limits}
Model-dependent limits were set for assumed IGMF strengths of B = 10$^{-14}$~G, 10$^{-15}$~G and 10$^{-16}$~G. As extended emission was not detected, magnetic fields in this range provide the most straightforward interpretation. In the case of non-detection, it is not possible to rule out magnetic field strengths in the regime in which the $e^{+}e^{-}$ pairs are isotropized and a pair halo forms without making strong assumptions about the source's history of activity. 

As discussed in the introduction, the projected sensitivity to the cascade emission hinges heavily on the source's intrinsic spectrum, and in particular to high energy cut-off. For VER J0521+211, 1ES 0229+200, and PG 1553+113, $E_{cut}$ is taken to be 10~TeV and the fraction of the total emission due to cascade emission is within VERITAS sensitivity (cascade fraction$>$5\%). For the other sources, the predicted cascade fraction is too small to warrant setting model-dependent limits.

Model predictions were obtained from a 3-dimensional semi-analytical cascade code~\cite{Weisgarber}, using the Gilmore EBL model~\cite{Gilmore2012} and assuming an IGMF coherence length of 1 Mpc. The blazar jet's Doppler factor was set to 10, and the viewing angle was set to $0^{\circ}$. The simulation took into account the source redshift and events were weighted to match the desired intrinsic spectrum. 

An ensemble of pseudo-experiments for the extended emission was generated by convolving the predicted extended emission $\theta^{2}$ distribution with the PSF distribution corresponding to the source of interest. The uncertainty on the point source width was incorporated by sampling a Gaussian distribution with a width reflecting the statistical and systematic errors on $w_{sim}$ (hence the use of pseudo-experiments rather than a single convolution). A second ensemble of pseudo-experiments at each cascade fraction $f_{c}$ was generated. The appropriate number of $\gamma$-ray events for each source was sampled according to [$(1 - f_{c})\times$ Direct emission + $f_{c}\times$ Extended emission].

\begin{figure}
\centerline{\includegraphics[width=0.42\textwidth]{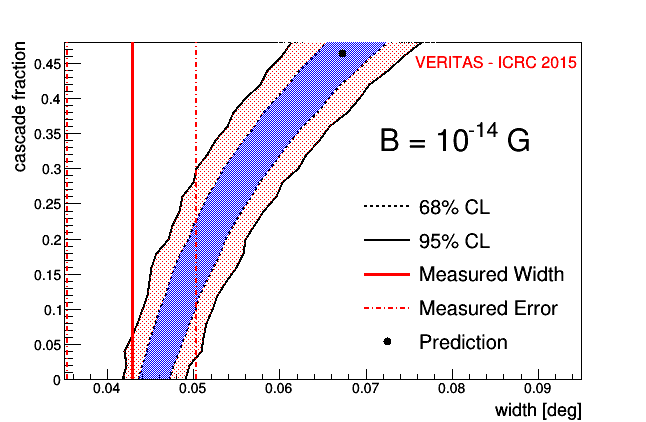}
\includegraphics[width=0.42\textwidth]{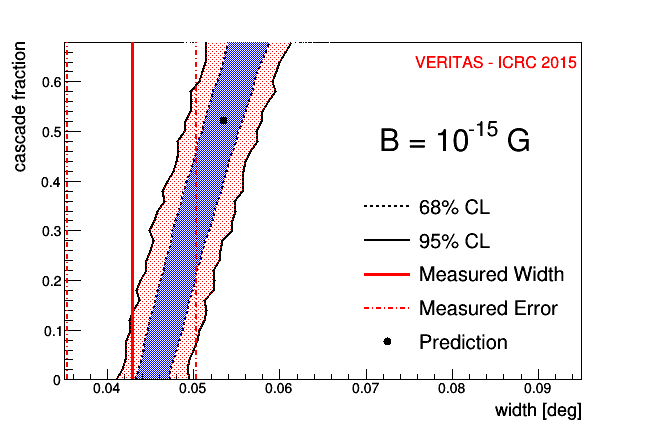}}
\centerline{\includegraphics[width=0.42\textwidth]{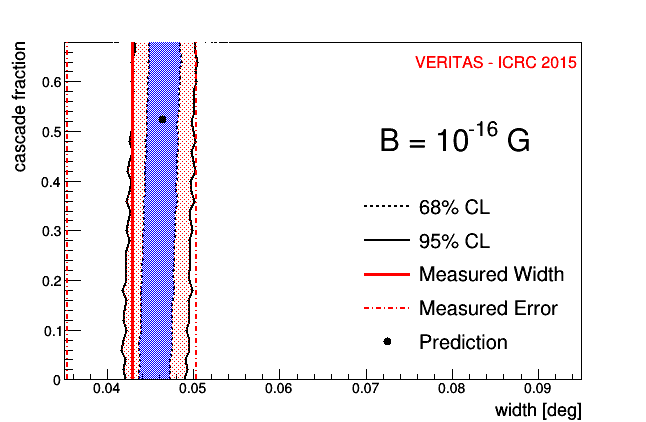}}
\caption{\small{Confidence intervals (68\% and 95\% CL) on the width of the $\theta^{2}$ distribution for a range of cascade fractions, using the source 1ES 0229+200 and assumed IGMF strengths B=10$^{-14}$, 10$^{-15}$, and 10$^{-16}$ G.}}\label{sigma_v_fc_1ES0229}
\end{figure}

Fig.~\ref{sigma_v_fc_1ES0229} shows the fitted width of pseudo-experiments for 1ES 0229+200 versus $f_{c}$, overlaid with the width measured in data. The uncertainties on the measured and simulated widths, added in quadrature, are considered in the extraction of upper limits on $f_{c}$. Upper limits were derived for PG 1553+113 and VER J0521+211 as well, however 1ES 0229+200 allows the strongest constraints on $f_{c}$ compared to the prediction. For B = 10$^{-16}$~G, no values of $f_{c}$ are excluded. For B = 10$^{-14}$~G, the cascade fraction is less than 0.18 at the 68\% confidence level (CL) and less than 0.35 at the 95\% CL. For B = 10$^{-15}$~G, the cascade fraction is less than 0.36 at the 68\% CL and less than 0.78 at the 95\% CL. Based on predicted values and the measured upper limits for the cascade fraction, IGMF strengths in the range (5.0--10.0)$\times 10^{-15}$~G are ruled out at the 95\% CL. The reader is reminded that this constraint only applies to the range of field strengths considered, and fields strong enough to produce pair halos have not been ruled out by non-detection.

The model-dependent limits show similar sensitivity to measurements from other instruments. In particular, H.E.S.S. ruled out an IGMF strength in the range (0.3--3)$\times$10$^{-15}$~G with slightly different model assumptions~\cite{HESS}.

\subsection{Model-independent limits}
Model-independent upper limits on the flux due to extended emission are set for all sources. The bulk of the beamed emission is expected to fall in the range $\theta^{2}$ = 0.0--0.01, thus the excess counts due to extended emission was calculated from the difference $\int \theta^{2}_{data} - \int \theta^{2}_{sim}$ with integration limits $\theta^{2}=0.01$ to $\theta^{2}=0.24$. An upper limit on the $\gamma$-ray rate due to extended emission was calculated using the method of Rolke~\cite{Rolke}. Translating the upper limit on the rate to a flux upper limit requires an assumption for the spectral index of the cascade emission. Two possible values were used, the assumed intrinsic index $\Gamma$, and a softer spectrum with spectral index ($\Gamma+1$). The resulting 99\% confidence level upper limits are shown Tab.~\ref{modelindlimits}.

\begin{table}
\centerline{
\begin{tabular}{ccc}
  & $99\%~CL(\Gamma)$ &  $99\%~CL(\Gamma+1)$  \\
\hline
\hline
& [$10^{-12}cm^{-2}TeV^{-1}s^{-1}$] & [$10^{-12}cm^{-2}TeV^{-1}s^{-1}$] \\
\hline
\hline
Mrk 421 & 0.62 & 0.50 \\
Mrk 501 & 2.69 & 2.16 \\
1ES 0229+200 & 0.64 & 0.58 \\
H 1426+428 & 1.23 & 1.18 \\ 
VER J0521+211 & 0.20 & 0.17 \\ 
1ES 1218+304 & 0.58 & 0.46 \\ 
PG 1553+113 & 1.83 & 1.48 \\ 
\hline
\hline
\end{tabular}
}
\caption{Model-independent limits on the flux from extended emission, assuming cascade emission with a spectral index matching the intrinsic spectral index $\Gamma$ and ($\Gamma$+1).}
\label{modelindlimits}
\end{table}

\section{Conclusions}
A search for source extension due to cascade emission broadened by the IGMF was performed using long VERITAS exposures on a number of hard-spectrum blazars. No indication of extended emission was observed. Model-dependent limits were set on the fraction of the total emission due to cascade emission, assuming IGMF strengths between 10$^{-16}$ and 10$^{-14}$~G. IGMF strengths in the range (5.0--10.0)$\times 10^{-15}$~G were excluded at the 95\% CL, assuming an IGMF coherence length of 1 Mpc. Model-independent limits were set on the flux due to extended emission, resulting in 99\% CL upper limits of 0.17--2.69$\times10^{-12}$~cm$^{-2}$~TeV$^{-1}$~s$^{-1}$ for an energy range between 160~GeV and 1~TeV.
\pagebreak
\\[2ex]
\textit{Acknowledgments} \\
\small{This research is supported by grants from the U.S. Department of Energy Office of Science, the U.S. National Science Foundation and the Smithsonian Institution, by NSERC in Canada, by Science Foundation Ireland (SFI 10/RFP/AST2748) and by STFC in the U.K. We acknowledge the excellent work of the technical support staff at the Fred Lawrence Whipple Observatory and at the collaborating institutions in the construction and operation of the instrument. The VERITAS Collaboration is grateful to Trevor Weekes for his seminal contributions and leadership in the field of VHE gamma-ray astrophysics, which made this study possible. We thank T. Weisgarber for providing model predictions. The funding for this work was provided by a Marie Curie Intra-European Fellowship.}

\end{document}